\documentclass[conference]{IEEEtran}
\IEEEoverridecommandlockouts
\usepackage{booktabs}
\usepackage[ruled,vlined,linesnumbered]{algorithm2e}
\usepackage{amssymb,amsmath,epsfig,latexsym,graphicx,bm}
\usepackage{epstopdf}
\usepackage{xcolor}
\usepackage{soul}
\usepackage{cite}
\usepackage{amsfonts}
\usepackage{amsmath}
\usepackage{amsmath,amssymb}
\usepackage{textcomp}
\usepackage{xcolor}
\usepackage{optidef}

\usepackage[dvipsnames]{xcolor}

\definecolor{amaranth}{rgb}{0.9, 0.17, 0.31}

\def\BibTeX{{\rm B\kern-.05em{\sc i\kern-.025em b}\kern-.08em
		T\kern-.1667em\lower.7ex\hbox{E}\kern-.125emX}}

\usepackage[font=footnotesize,caption=false]{subfig}

\title{Toward ISAC-empowered subnetworks: Cooperative localization and iterative node selection}
\author{\IEEEauthorblockN{Mostafa Nozari\IEEEauthorrefmark{1}, Israel Leyva-Mayorga\IEEEauthorrefmark{1}, Fabio Saggese\IEEEauthorrefmark{2}, and Gilberto Berardinelli\IEEEauthorrefmark{1}} 
\IEEEauthorblockA{\IEEEauthorrefmark{1}Department of Electronic Systems, Aalborg University, Denmark (\{mnozari, ilm, gb\}@es.aau.dk)} 
\IEEEauthorblockA{\IEEEauthorrefmark{2}Department of Information Engineering, University of Pisa, Italy (fabio.saggese@ing.unipi.it).}
\thanks{This work is supported by Independent Research Fund Denmark, grant no. 3105-00077B. The work of F. Saggese is supported by Horizon Europe MSCA Postdoctoral Fellowships with Grant~101204088.}
}

\begin{document}
\maketitle
\bstctlcite{IEEEexample:BSTcontrol}
\begin{abstract}
This paper tackles  the sensing–communication trade-off in integrated sensing and communication (ISAC)-empowered subnetworks for mono-static target localization. We 
propose a low-complexity iterative node selection algorithm that exploits the spatial diversity of subnetwork deployments and dynamically refines the set of sensing subnetworks to maximize localization accuracy under tight resource constraints. Simulation results show that our method achieves sub-7\,cm accuracy in additive white Gaussian noise (AWGN) channels within only three iterations—yielding over 97\% improvement compared to the best-performing benchmark under the same sensing budget. We further demonstrate that  increasing spatial diversity through additional antennas and subnetworks enhances sensing robustness, especially in fading channels. Finally, we quantify the sensing-communication trade-off 
showing that reducing sensing iterations and the number of sensing subnetworks improves throughput at the cost of reduced localization precision. 
\end{abstract}

\section{Introduction}
In the 6th Generation (6G) `network of networks vision', in-X subnetworks are expected to offer ultra-local, high-performance wireless connectivity within entities such as robots, vehicles, and production units~\cite{berardinelli2021extreme}. In-X subnetworks can become spontaneously dense, particularly in industrial settings with many robots, and may support mission-critical tasks. They must therefore offer standalone capabilities, like ad hoc networks, while retaining the ability to connect to the  6G global network to tap into functions such as resource coordination, traffic management, monitoring, and authentication.

Concurrently, 6G radio is expected to support Integrated Sensing and Communication (ISAC), enabling shared infrastructure for data transmission and environmental perception~\cite{lu2024integrated}. This integration allows in-X subnetworks to evolve into perceptive networks—supporting both local communication and collaborative sensing tasks such as localization,  mapping, and motion classification.The spatial diversity of dense subnetwork deployments also provides multiple environmental viewpoints, enhancing sensing accuracy and robustness~\cite{Meng2025cooperative}.

Recent research on in-X subnetworks has focused on managing interference in dense deployments through techniques such as power control and subband allocation~\cite{srinivasan2024multi, 
hakimi2025resilient}. Both distributed strategies and centralized approaches, where decisions are coordinated by an edge server (ES), have been investigated. However, the potential of leveraging in-X subnetworks for sensing purposes remains unexplored. 


Collaborative sensing has gained significant attention in recent ISAC research, focusing primarily on interference mitigation, beamforming design, power and resource allocation \cite{jiang2023collaborative,meng2024network}.
However, effectively leveraging the geometric diversity of spatially distributed nodes through intelligent node selection for sensing remains an open and critical challenge \cite{Li2025cooperative}.


This paper presents an initial study on integrating ISAC into in-X subnetworks, with a focus on target localization in dense industrial environments.
Although dense in-X subnetworks offer substantial geometric diversity for accurate  sensing, distributing sensing tasks across a large number of subnetworks may cause unacceptable delays for the localization task and degrade communication performance due to the intensive use of resources. 
Therefore, a rapid selection of subnetwork subsets with optimal visibility to the target is essential for efficient localization under time and bandwidth constraints.

To this end, we propose a subnetwork subset selection technique based on a surrogate Weighted Geometric Dilution of Precision (WGDOP) criterion~\cite{chen2013calculation}, enabling fast and efficient selection of the most informative sensing agents.
The main contributions of the paper are the following:
\begin{itemize}
    \item We introduce a frame structure and general system operation for incorporating sensing functionalities into communication-native in-X subnetworks.
    \item We propose an iterative algorithm for selecting a subset of sensing subnetworks with limited delay and resource usage. 
    To enhance localization robustness, we apply a weighted least squares (WLS) fusion that accounts for measurement reliability across the selected subnetworks.
    
    \item We derive the Cramér–Rao lower bound (CRLB) for ranging and introduce a practical approximation, 
    the empirical CRLB (eCRLB), which captures link quality. This metric is integrated into the WGDOP criterion, enabling subnetwork selection based on both geometry and link quality information.  
    \item We evaluate  the proposed solution through comprehensive simulations, assessing the localization accuracy and the associated latency–communication trade-offs. 

\end{itemize}

To support low-capability devices and limited bandwidth, our method relies on received signal strength (RSS). Nonetheless, it is general and adaptable to other ranging methods, such as Time of Arrival (ToA) or Angle of Arrival (AoA).

\section{System Model} \label{SystemModel}
\begin{figure}[t]
\centering
\includegraphics[width=0.80\columnwidth]{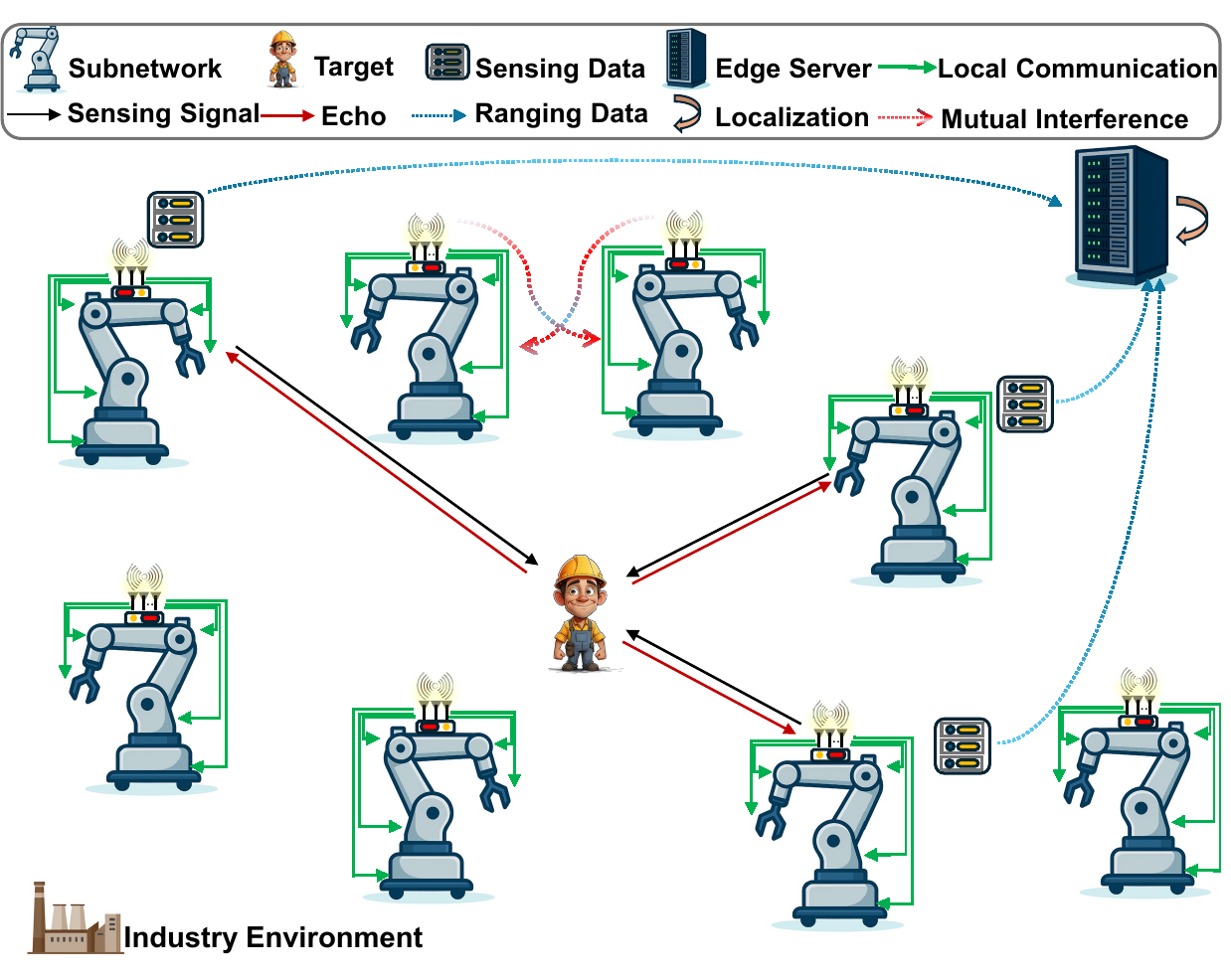}
\caption{IIoT environment where $K=3$ out of $N = 8$ ISAC-empowered subnetworks employ monostatic sensing to collaboratively localize a target, while all subnetworks support local communication.}
\label{SYYYSSIIMO}
\vspace{-5mm}
\end{figure}

We consider a 6G 
network deployed in a dense Industrial Internet of Things (IIoT) environment comprising an ES and \(N\) subnetworks that employ orthogonal frequency-division multiple access (OFDMA) for communication. 
Each subnetwork is equipped with an access point (AP) that supports wireless connectivity to $U$ internal devices using the shared OFDMA resources and to the ES using a dedicated out-of-band control channel,
as illustrated in Fig.~\ref{SYYYSSIIMO}. 
In addition, the subnetworks implement a distributed mono-static sensing framework to localize a single target, thereby enhancing situational awareness and operational safety. The network is modeled in a two-dimensional Cartesian coordinate system, where the location of the subnetwork AP \( n \) is denoted by \( \mathbf{s}_n = [x[n],\, y[n]] \), and the target location by \( \mathbf{q} = [x_t,\, y_t] \).  Each subnetwork's AP is equipped with a single-antenna for downlink (DL) communication and sensing signal transmission, while an \( M \)-element uniform linear array (ULA) of receive antennas is used for echo reception during sensing.

\subsection{Frame Structure}
The subnetworks operate over a discretized time–frequency grid, following the OFDMA principle. 
Therefore,
each frame of duration $T$ seconds is divided into \( F \) time slots of equal length, and 
\( N_{\mathrm{RB}} \) orthogonal resource blocks (RBs). An RB is the fundamental time-frequency unit for resource allocation, which 
consists of $R$ orthogonal resource elements (RE), 
spanning a fixed number of subcarriers and OFDM symbols.

\begin{figure}[t]
    \centering
    \includegraphics[width=0.9\columnwidth]{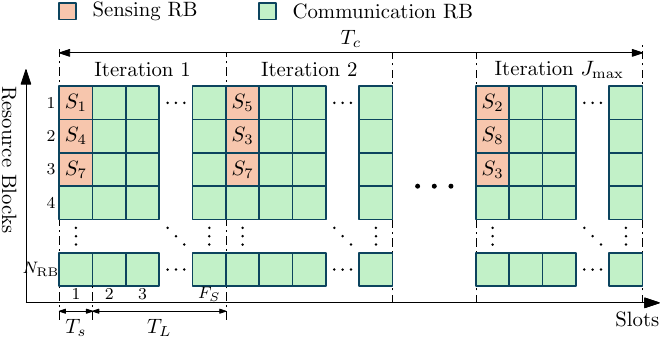}
    \caption{Example of the proposed algorithm frame structure. Communication RBs are used by all subnetworks, whereas the $\rho = K = 3$ sensing RBs per iteration $j$ are reserved the subnetworks in $\mathcal{S}^{(j)}$ to avoid interference.} 
    \label{fig:Fr} 
\end{figure}

The 6G network performs global time-frequency synchronization and allocation across all subnetworks by dividing the RBs into sensing or communication. 
To avoid mutual interference during 
sensing, each sensing RB is assigned to a specific subnetwork (see Fig. \ref{fig:Fr}). 
Conversely, since communication among APs and internal devices is short range, communication RBs are used by all subnetworks simultaneously. Although subnetworks share communication resources non-orthogonally, each AP utilizes orthogonal frequency multiplexing to simultaneously serve its \( U \) internal users.

Without loss of generality, we assume the sensing process consists of one or more iterations, with total duration upper bounded by time budget \( T_c \), assumed small enough that \( \mathbf{q} \) and \( \mathbf{s}_n, \, \forall n \), remain static. Each iteration comprises a \emph{sensing period} of duration \( T_s \), followed by a \emph{localization-feedback period} of duration \( T_L \), and spans \( F_S \) time slots. The number of RBs reserved for sensing in each iteration is denoted by \( \rho \). Accordingly, maximum number of iterations is bounded by
\begin{equation}
J_{\max} \leq \left\lfloor\frac{T_c}{T_s+T_L}\right\rfloor. 
\label{eq:frame_constraint}
\end{equation}
Let us consider the $j^\text{th}$ sensing iteration $j\in\{1,2,\dotsc, J_{\max}\}$. 
Within the sensing period, a subset of subnetworks  \(\mathcal{S}^{(j)} \subseteq \{1, 2, \dots, N\}\) 
performs mono-static sensing.  During the localization-feedback period, the selected subnetworks report their measurements to the ES via the out-of-band control channel. Upon receiving the measurements, the ES generates a single estimate for the target location and, if needed, sends feedback to the subnetworks with the configuration for the next sensing iteration. 
In practice, the ES may adaptively configure 
\( T_L\) 
based on traffic load, 
the  subset size \( |\mathcal{S}^{(j)}| \),
and quality-of-service (QoS) demands, enabling flexible trade-offs between sensing accuracy, latency, and communication throughput. 

\subsection{Sensing Signal Model}
The considered mono-static sensing framework relies on transmitting pilot signals in specific 
REs and measuring the 
RSS to localize a single target. 
To enable coordinated sensing, in each sensing iteration, the ES selects a subset of \( K \leq N \) subnetworks and assigns $\rho / K$ orthogonal RBs to each of them. 
Let  \( \mathcal{X}_n=\{x_n[r]\}_{r=1}^{R} \) denote the pilot signal transmitted by subnetwork \( n \) at a given sensing iteration. The pilot signal consists of $R$ individual pilot symbols transmitted at a set of REs \(\{r\} \), given by \( x_n[r] = \sqrt{p_n} \, u[r] \), where \( u[r] \in \mathbb{C} \) is a unit-power symbol satisfying \( \mathbb{E}[|u[r]|^2] = 1 \), and \( p_n \) denotes the transmit power. By assuming perfect self-interference cancellation capabilities at the APs~\cite{liu2023snr}, the reflected signal received at antenna \( m \) of subnetwork $n$ in RE $r$
is given as
\begin{equation}
y_n^{(m)}[r] = h_n^{(m)}[r] x_n[r] + z_n^{(m)}[r],
\label{eq:scalar_model}
\end{equation}
where $z_n^{(m)}[r] \sim \mathcal{N}(0, \sigma^2_n)$ is the additive white Gaussian noise (AWGN), and \( h_n^{(m)}[r] \in \mathbb{C} \) is the target response coefficient in RE \( r \), at antenna \( m \) of subnetwork \( n \), encompassing the direct impinging signal and the reflection from the target. Assuming half-wavelength antenna spacing, i.e., $\frac{\lambda}{2}$, where $\lambda$ is the carrier wavelength, the reflected signal reaches each antenna through independently faded paths. Hence, the received signal power at AP \( n \) after transmitting $R$ pilot symbols is estimated as
\begin{equation} 
    \hat{P}^{tot}_{n} = \left(\frac{1}{MR} \sum_{r=1}^{R}\sum_{m=1}^{M}\left|y^{(m)}_n[r]\right|^2\right) - \sigma^2_n. 
    \label{div}
\end{equation}
If $\sigma^2_n$ is perfectly known, 
$\hat{P}^{tot}_{n}$ serves as an unbiased estimator of the RSS, 
derived from the radar range equation
\begin{equation}
\mathbb{E}\left[\hat{P}^{\mathrm{tot}}_n\right]= p_{n} G_t G_r \sigma_{\mathrm{rcs}} \cdot \frac{\lambda^2}{(4\pi)^3 d_n^4},
\end{equation}
where \( G_t \) and \( G_r \) represent the transmit and receive antenna gains, \( \sigma_{\mathrm{rcs}} \) is the radar cross-section (RCS) of the target, 
and \( d_n \) denotes the true distance between subnetwork \( n \) and the target. 
Consequently, an estimate for $d_n$ can be formulated as
\begin{equation}
\hat{d}_n = \left( \frac{p_n G_t G_r \sigma_{\mathrm{rcs}} \cdot \frac{\lambda^2}{(4\pi)^3}}{\hat{P}^{\mathrm{tot}}_n} \right)^{\frac{1}{4}}.
\label{RES}
\end{equation}

\subsection{Communication Signal Model}
\label{CMD}
DL communication takes place simultaneously in each subnetwork  \(n\) from its AP to a set of users \( \mathcal{U}_n = \{u_n(i)\}_{i=1}^U \) located at different distances from the AP.
Hence, the DL SINR at user \( u_n(i) \) served by AP \( n \) is 
\begin{equation}
\mathrm{SINR}_{n,u_n(i)} =
\frac{p_n\,|h_{n,u_n(i)}|^2}{\sigma_{u_n(i)}^2 + \sum_{j\neq n} p_j\,|h_{j,u_n(i)}|^2},
\label{eq:sinr}
\end{equation}
where 
\( h_{n,u_n(i)} \) denotes the complex channel gain, including pathloss, from AP \( n \) to user \( u_n(i) \), \( h_{j,u_n(i)} \) is the interference from AP \( j \ne n \), and \( \sigma_{u_n(i)}^2 \) is the noise power at \( u_n(i) \).
Hence, the ideal DL network sum-rate in communication RBs is 
\begin{equation}
R_0 = \sum_{n=1}^{N} \sum_{i=1}^{U} B \log_2\left(1 + \mathrm{SINR}_{n,u_n(i)}\right),
\label{eq:sumrate}
\end{equation}
where  \( B \) denotes the available bandwidth per user.

Having $\rho$ out of the $F_S N_\mathrm{RB}$ RBs being reserved for sensing in each sensing iteration,
the effective sum-rate becomes 
\begin{equation}
R_{\mathrm{eff}} = R_0\left(1 - \frac{\rho}{F_S\,N_{\mathrm{RB}}}\right).
\label{eq:rsense}
\end{equation}
This degradation causes a relative throughput loss of \( 
\frac{\rho}{F_S\,N_{\mathrm{RB}}} \), which scales linearly with \(\rho \), highlighting the trade-off between sensing and communication in ISAC design.

\section{Cooperative Localization: Iterative Node Selection and Refinement Approach}
\label{OS}
This section presents the proposed cooperative localization and iterative node selection approach. 
At each sensing iteration, 
a subset of subnetworks performs ranging using~\eqref{RES}, and forwards their measurements to the ES, which applies data fusion to obtain an estimate of the target location, denoted as
\( \hat{\mathbf{q}}\).
In the next iteration, 
the ES selects a new subset based on the geometric diversity of subnetworks relative to \( \hat{\mathbf{q}} \),
and their channel quality.
Geometric diversity reduces the ambiguity region formed by the intersection of annular range estimates, while high channel quality ensures reliable range estimates. 

To this end, optimization problem \textbf{P1} guides the ES in selecting a fixed-cardinality subset \( \mathcal{S} \subseteq \{1, 2, \dots, N\} \) with \( |\mathcal{S}| = K 
\geq 3 \) to ensure reliable RSS-based localization given an estimated location $\hat{\mathbf{q}}$. 
Once the optimal subset is identified, problem \textbf{P2} estimates a new target's position within the ambiguity region determined 
by \textbf{P1}. The process is repeated until convergence, as detailed in Sec.~\ref{sec:procedure}. Both problems 
are formally described next for a given sensing iteration.

\begin{subequations}\label{prob:P1}
\begin{align}
\textbf{P1:}\quad &
\mathcal{S}^{\star} = \;\arg\min_{\mathcal{S} \subseteq \{1,\dots,N\}} \;\text{Area} \left( \bigcap_{n \in \mathcal{S}} \mathcal{A}_n\right)
\label{prob:P1_obj}\\[0.4em]
\text{s.t.}\quad
&\mathcal{A}_n =
\left\{
\mathbf{q} \;\big|\; \lVert \mathbf{q} - \mathbf{s}_n\rVert^2_2 \in [d_{1n}, d_{2n}]
\right\}, \quad \forall n \in \mathcal{S}^{}
\label{prob:P1_annulus}\\[0.4em]
&\bigcap_{n \in \mathcal{S}^{}} \mathcal{A}_n \neq \emptyset
\label{prob:P1_nonempty}\\[0.4em]
&d_{2n} \ge d_{1n} > 0,\quad \forall n \in \mathcal{S}^{}
\label{prob:P1_bounds}\\[0.4em]
&\exists\,i_0\in\mathcal{S}^{}:\;
\operatorname{rank}\Bigl[\;\mathbf{s}_j - \mathbf{s}_{i_0}\Bigr]_{\,j\in\mathcal{S}^{}\setminus\{i_0\}} = 2.
\label{prob:P1_nocol}
\end{align}
\end{subequations}
The optimization problem \textbf{P1} seeks for the subset of sensing subnetworks leading to the smallest overlap of all annular regions \(\{\mathcal{A}_n\}\), thereby localizing the target with maximum precision. The parameters \(d_{1n}\) and \(d_{2n}\) in \eqref{prob:P1_annulus} define the inner and outer range bounds, modeling the range measurement uncertainty. 
The inner and outer radii of the annulus around each estimated distance are defined as 
\( d_{1n} = \widetilde{d}_n - \sqrt{ \widehat{\mathrm{CRLB}}_{n \leftarrow \hat{\mathbf{q}}} } \), and    
\( d_{2n} = \widetilde{d}_n + \sqrt{ \widehat{\mathrm{CRLB}}_{n \leftarrow \hat{\mathbf{q}}} } \),
 respectively,
where \( \widetilde{d}_n \) 
denotes the range approximation, computed using the current estimated target location 
i.e.,
$\widetilde{d}_n = \|\hat{\mathbf{q}}^{} - \mathbf{s}_n\|_2$,
and \( \widehat{\mathrm{CRLB}}_{n \leftarrow \hat{\mathbf{q}}} \) denotes the corresponding empirical variance bound, derived in the Appendix.
Constraint \eqref{prob:P1_nonempty} ensures that annulus intersection is nonempty—i.e., at least one feasible target location exists. Constraints \eqref{prob:P1_bounds} 
  guarantee valid annuli and constraint \eqref{prob:P1_nocol} excludes collinear subnetworks configurations. 
While \textbf{P1} depends on a target location estimate, we proceed by solving \textbf{P1} assuming this estimate is available, as detailed in~\ref{sec:procedure}. 

Directly verifying feasibility in \textbf{P1} by enumerating intersections across all \(\binom{N}{K}\) possible subsets is computationally prohibitive for large networks, since each annulus 
\(\mathcal{A}_n^{}\)
is nonconvex and lacks a closed-form intersection test. To circumvent this issue, we introduce a surrogate criterion based on the WGDOP metric. WGDOP 
captures both subnetwork geometry and measurement accuracy, admits a closed-form computation via simple matrix operations, and has an underlying convex structure, ensuring a unique global optimum. It is given by 
\begin{equation}
  \mathrm{WGDOP}(\mathcal{S}^{})
  = \sqrt{\operatorname{Tr}\!\Bigl[\bigl(\mathbf{H}_{\mathcal{S}^{}}^\top
                                   \,\mathbf{W}_{\mathcal{S}^{}}
                                   \,\mathbf{H}_{\mathcal{S}^{}}\bigr)^{-1}\Bigr]},
  \label{eq:wgdop}
\end{equation}
where $\mathrm{Tr}(\cdot)$ is the trace operation, $\mathbf{H}_{\mathcal{S}}\in\mathbb{R}^{|\mathcal{S}| \times 2}$ is the geometry matrix, with rows
\begin{equation}
 \left[\mathbf{H}_{\mathcal{S}^{}}\right]_{n,:}
 =
\frac{\hat{\mathbf{q}}^{} - \mathbf{s}_n}
{\lVert \hat{\mathbf{q}}^{} - \mathbf{s}_n \rVert_2}, 
\quad n \in \mathcal{S}, 
\label{eq:jacobian_row}
\end{equation}
%
and $\mathbf{W}_{\mathcal{S}^{}}\in\mathbb{R}^{|\mathcal{S}| \times |\mathcal{S}|}$ is an eCRLB-based diagonal weight matrix
\begin{equation}
\mathbf{W}_{\mathcal{S}^{}} = 
\mathrm{diag} \left[ \left(\widehat{\mathrm{CRLB}}_{n \leftarrow \hat{\mathbf{q}}}\right)^{-1} \right]_{n\in\mathcal{S}}.
\label{eq:weight_matrix}
\end{equation}

Thus, we reformulate \textbf{P1} as a WGDOP minimization problem that jointly accounts for geometric diversity via~\eqref{eq:jacobian_row} and sensing channel quality via~\eqref{eq:weight_matrix}, thereby promoting favorable subnetwork configurations for sensing; i.e.,
%
\begin{subequations}\label{prob:P2}
\begin{align}
\textbf{P1.1:}\quad 
&\mathcal{S}^{\star }
   =\arg\min_{\mathcal{S}^{}\subseteq\{1,\dots,N\}}\;\mathrm{WGDOP}(\mathcal{S}^{})
\label{prob:P2_obj}\\[0.4em]
\text{s.t.}\quad 
&
\det\!\bigl(\mathbf{H}_{\mathcal{S}^{}}^\top\,\mathbf{W}_{\mathcal{S}^{}}\,\mathbf{H}_{\mathcal{S}^{}}\bigr)>0,
\label{prob:P2_fullrank}
\end{align}
\end{subequations}
where~\eqref{prob:P2_fullrank} implies 
$\mathrm{rank}(\mathbf{H}_{\mathcal{S}^{}})=2$.
Although the underlying cost function in \textbf{P1.1} is convex, the optimization requires evaluating all possible subnetwork subsets of fixed cardinality, making the overall problem combinatorial. 
Therefore, we obtain the optimal subnetwork subset $\mathcal{S}^{\star}$ via exhaustive search, which is feasible for moderate values of \( N \) and \( K \).

The subnetworks of the optimal subset $\mathcal{S}^{\star}$ obtained from \textbf{P1.1} perform range estimations $\hat{d}_n$ through~\eqref{RES}. Then, the target location problem \textbf{P2} can be formulated as
%
\begin{equation}
\begin{aligned}
\textbf{P2:}\,\,
[x_t^\star, y_t^\star] = 
&\arg\hspace{-4mm}\min_{\mathbf{q} \in \bigcap_{n \in \mathcal{S}^{\star}} \mathcal{A}_n} \hspace{-1mm}
\sum_{n \in \mathcal{S}^{\star}}  \hspace{-1.2mm} \frac{(\lVert\mathbf{q} - \mathbf{s}_n\rVert_2- \hat{d}_n)^2}{ \widetilde{\mathrm{CRLB}}_{n}}.  
\end{aligned}
\label{eq:wls_opt}
\end{equation}
 \textbf{P2} is solved using an iterative Gauss--Newton weighted least squares (WLS) method~\cite{bjorck2024numerical}, which incorporates reliability-aware weights to account for the quality of individual range measurements prior to data fusion. The weight for each subnetwork \( n \) is given by \( \widetilde{\mathrm{CRLB}}_{n} \), which has the same form as the theoretical CRLB derived in Appendix, but the true range is replaced by the measured range \( \hat{d}_n \) computed using~\eqref{RES}.

Let \( \mathbf{q}^{(i)} = [x_t^{(i)}, y_t^{(i)}] \) denote the target position estimate at the \( i^\text{th} \) WLS iteration. The residual vector  at the \( i^\text{th} \) WLS iteration, denoted \( \mathbf{e}^{(i)} \), is defined as
\begin{equation}
\mathbf{e}^{(i)} = \left[ \hat{d}_n - \lVert \mathbf{q}^{(i)} - \mathbf{s}_n \rVert_2 \right]_{n \in \mathcal{S}^{\star}}.
\label{eq:residual_vector}
\end{equation}
The Gauss-Newton update step is then computed as~\cite{bjorck2024numerical}
\begin{equation}
  \Delta\mathbf{q}^{(i)}
  = \bigl(\mathbf{H}_{\mathcal{S}^{\star }}^{(i)\top}\mathbf{W}_{\mathcal{S}^{\star }}\mathbf{H}_{\mathcal{S}^{\star}}^{(i)}\bigr)^{-1}
    \mathbf{H}_{\mathcal{S}^{\star}}^{(i)\top}\mathbf{W}_{\mathcal{S}^{\star}}\,\mathbf{e}^{(i)},
  \label{eq:delta_x}
\end{equation}
where \( \mathbf{H}_{\mathcal{S}^{\star}}^{(i)} \in \mathbb{R}^{|\mathcal{S}| \times 2} \) is the Jacobian matrix defined in \eqref{eq:jacobian_row}, evaluated at the target location 
\( \mathbf{q}^{(i)} \), and \( \mathbf{W}_{\mathcal{S}^{\star}} \) 
is the matrix in~\eqref{eq:weight_matrix}, whose diagonal elements are given by \( 1 / \widetilde{\mathrm{CRLB}}_{n} \).
The target position estimate is then updated as
\begin{equation}
  \mathbf{q}^{(i+1)} = \mathbf{q}^{(i)} + \Delta\mathbf{q}^{(i)}.
  \label{eq:update}
\end{equation}
The process starts from an initial arbitrary position guess $\mathbf{q}^{(1)}$ and continues iteratively until convergence. Convergence is declared when the two successive estimates satisfies \( \|\mathbf{q}^{(i+1)} - \mathbf{q}^{(i)}\| \leq \varepsilon \), for some small predefined threshold \( \varepsilon > 0 \), or when the maximum number of WLS iterations is reached. The value of \( \mathbf{q} \) at WLS convergence corresponds to 
\( \mathbf[x_t^\star, y_t^\star] \). 

\subsection{Algorithm Procedure}
\label{sec:procedure}
We developed an algorithm that solves \textbf{P1.1} and \textbf{P2} via iterative steps. The procedure is detailed in Algorithm~\ref{alg1}. 

At each iteration, the algorithm estimates the target position and updates the subset of sensing subnetworks for the next iteration based on the WGDOP criterion.
At the initial iteration, \( j = 0 \), the ES randomly selects a subset of \( K \) subnetworks, i.e., \( |\mathcal{S}^{(0)}| = K \), to perform sensing. After collecting the range estimates from these subnetworks, the ES executes the inner WLS routine to compute the initial estimate of the target location. 
Using this estimate, the ES evaluates the WGDOP metric 
through ~\eqref{eq:wgdop} 
for all possible ${N \choose K}$ combinations of subsets. 
Among these, the subset with the lowest WGDOP, $\mathcal{S}^{\star(1)}$, is selected for sensing in the next iteration.
%
%

The process terminates when 
the selected subnetwork subset remains unchanged across consecutive iterations, 
or when the iteration limit is reached.
The value of \( \hat{\mathbf{q}} \) at convergence is taken as the final target location estimate, denoted by \( \hat{\mathbf{q}}^\star \).

\medskip

\medskip

\begin{algorithm}[t]
\footnotesize
\caption{Cooperative Localization: Iterative Node Selection and Refinement}\label{alg:wgdop}
\label{alg1}
\LinesNumbered
\KwIn{Subnetwork positions $\{\mathbf{s}_n\}_{n=1}^N$, subset size $K$, max. no. of outer- and WLS iterations $J_{\max}$ and $I_{\max}$, threshold $\varepsilon$.} 
\KwOut{Estimated location $\hat{\mathbf{q}}^\star$, selected subset $\mathcal{S}^\star$.}
\BlankLine
\textbf{Initialize:}\  
Pick $\mathcal{S}^{(0)}\subset\{1,\dots,N\}$ 
 with $|\mathcal{S}^{(0)}|=k$; 
set $j\gets0$;

\While{ 
$j < J_{\max}$}{%
  Compute \( \hat{d}_n \) for all \( n \in \mathcal{S}^{(j)} \) via~\eqref{RES}\;
   Initialize $\mathbf{q}^{(1)}$\;
  \For{$i \gets 1$ \KwTo $I_{\max}$}{%
     Solve $\Delta\mathbf{q}^{(i)}$ via \eqref{eq:delta_x}\;
     Update $\mathbf{q}^{(i+1)} \gets \mathbf{q}^{(i)} + \Delta\mathbf{q}^{(i)}$ \eqref{eq:update}\;
       \If{ $\|\mathbf{q}^{(i+1)} - \mathbf{q}^{(i)}\| \leq \varepsilon$ }{%
    \textbf{break}\
  }
  }
   Set $\hat{\mathbf{q}} \gets \mathbf{q}^{(i)}$; 
  $j \gets j+1$\;

   Evaluate $\mathrm{WGDOP}(\mathcal{S}^{(j)})$ for all $|\mathcal{S}^{(j)}| = K$ via \eqref{eq:wgdop}\;
   $\mathcal{S}^{\star(j)} \gets \arg\min \mathrm{WGDOP}(\mathcal{S}^{(j)})$\;

  \If{$\mathcal{S}^{\star(j)} = \mathcal{S}^{\star(j-1)}$}{%
  \textbf{break};\quad The algorithm has reached convergence
  }
}
\Return $\mathcal{S}^{\star} \gets \mathcal{S}^{\star(j)}\;,$ 
$\hat{\mathbf{q}}^\star \gets \hat{\mathbf{q}}$ 
\end{algorithm}

\section{Numerical Results and Discussion} \label{Numerical}

In this section, we evaluate the performance of the proposed algorithm.
All results are averaged over \( 10^5 \) independent Monte Carlo simulations. 
The key simulation parameters are summarized in Table~\ref{tab:sim_params}. 
The scenario represents an industrial setting with $N$ subnetworks  deployed uniformly at random in a \(200\,\mathrm{m} \times 200\,\mathrm{m}\) area, and a single human target with an RCS of \(0\,\mathrm{dBsm}\)~\cite{luo2024channel}.
The operating wavelength correspond to a carrier frequency of \(10\,\text{GHz}\).
The frame follows 5G NR specification with numerology zero. It consists of a $T = 10$ ms frame divided into \(F = 10\) slots, each lasting for $1$ ms and comprising \(14\) OFDM symbols. Each RB spans \(12\) subcarriers with $15$ kHz spacing. Thus, the bandwidth of $B_T=20$ MHz corresponds to $N_\text{RB} = 106$.  The sensing duration within each frame is \(T_s = 1\,\text{ms}\). 
Thermal noise power spectral density is $-174\,\mathrm{dBm/Hz}$. The sensing channels have expression $h_n^{(m)}[r] = \sqrt{\sigma_\mathrm{rcs} \frac{G_t G_r \lambda^2}{(4 \pi)^3 d_n^4}} \alpha_n^{(m)}[r]$, where $\alpha_n^{(m)} [r] = 1$ for AWGN, $ \alpha_n^{(m)}[r] \sim \mathcal{CN}(0,1)$ for Rayleigh, and $|\alpha_n^{(m)}[r]|$ follows a Rice distribution with a \( \kappa \)-factor of 7 and unit power for the Rician fading. The communication channels are modeled as AWGN, assuming a short-range line-of-sight (LoS) link, with the $U$ users of each subnetwork being uniformly distributed in an annular region centered at their respective APs, with radii from 2 to 6 m.

\begin{table}[t]
\footnotesize
\centering
\caption{Simulation Parameters}
\label{tab:sim_params}
\renewcommand{\arraystretch}{1.2}
\begin{tabular}{@{}llp{2.2cm}@{}}

\toprule
\textbf{Notation} & \textbf{Definition} & \textbf{Value} \\
\midrule
$N$ & Number of subnetworks & 40 \\
$U$ & Users per subnetwork & 5 \\
$K$ & Cardinality of $\mathcal{S}^{(j)}$ & 3 \\
$\lambda$ & Carrier wavelength & 0.03\,m \\
$p_n$ & Transmit power & 23\,dBm \\
$\sigma_{\mathrm{rcs}}$ & Radar cross section & 0\,dBsm~\cite{luo2024channel} \\
$B_T$ & Bandwidth  & 20\,MHz \\
$T$ & Frame duration & 10\,ms \\
$F$ & Number of time slots per frame & 10 \\
$T_s$ & Sensing duration & 1\,ms \\
$G_t, G_r$ & Transmit and receive antenna gain & 0\,dBi \\
\bottomrule
\end{tabular}
\vspace{-0.5cm}
\end{table}

\begin{figure}[t]
    \centering
    \includegraphics[width=0.50\linewidth]{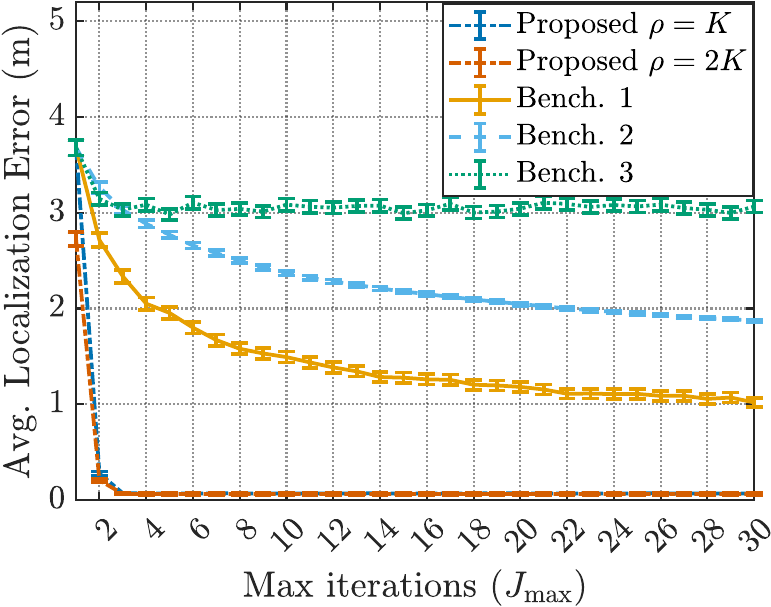}
    \caption{Average localization error and $90$\% CI vs. algorithm iterations over an AWGN Channel.}
    \label{fig:f1} 
    \vspace{-0.5cm}
\end{figure}

We consider two configurations for the proposed method, with different numbers of sensing RBs per iteration, which are equally divided among sensing subnetworks. The first configuration uses $\rho=K$, while the second allocates $\rho = 2K$ RBs per iteration. The performance of both configurations is evaluated against three benchmarks.  \emph{Benchmark 1}   selects a subset of subnetworks uniformly at random and performs a single sensing iteration, using $J_\text{max}\rho$ RBs for fairness. \emph{Benchmark 2} is an iterative approach that randomly selects subnetworks at each iteration, allocates $\rho=K$ RBs per iteration, and computes the final estimate by averaging all intermediate ones. Finally, \emph{Benchmark 3} follows the same procedure as Benchmark 2 but initializes the WLS solver with the previous estimate to accelerate convergence.
Fig.~\ref{fig:f1} illustrates the average absolute localization error and its 90\% confidence interval (CI)  
under an AWGN channel as a function of the maximum number of iterations $J_\text{max}$ for the two configurations of the proposed method and the three benchmarks. 
For $J_\text{max}= 1$, the proposed method with $\rho=K$  and the three benchmarks yield a similar localization error of approx. 3.6\,m, as no optimization has yet been performed. In contrast, the proposed method with $\rho=2K$, 
achieves a lower initial error of around 2.7\,m. 
Although the benchmarks exhibit some error reduction as $J_\text{max}$ increases, with Benchmark~1 consistently outperforming the rest, none of them achieve errors below $1$~m. Conversely, 
when $J_\text{max} = 2$, the proposed method achieves a localization error below 27\,cm for $\rho=K$ and 20\, cm for $\rho=2K$.
For $J_\text{max}= 3$, both configurations of the proposed method 
reach localization errors of around 6.7\,cm and 6.1\,cm, respectively, and additional iterations have a negligible impact. 
These errors are more than $97$\% lower than those achieved by the benchmarks with the same iterations. 

Both $\rho=K$ and $\rho=2K$ configurations achieve comparable localization accuracy for $J_\text{max}\ge 3$ and meet the accuracy target of 10\,cm for indoor environments~\cite{zafari2019survey}.  We therefore adopt the $\rho=K$ configuration—since it requires only half the sensing RBs—for all subsequent evaluations.


\begin{figure}[t]
  \centering
  \subfloat[Average Localization Error]{%
    \includegraphics[width=0.495\linewidth]{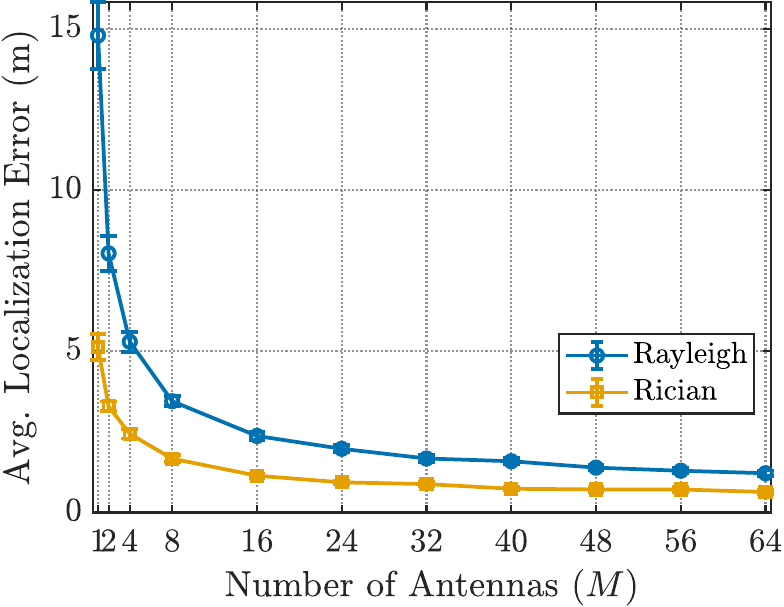}%
    \label{fig3:avg_error}}
    \hfil
  \subfloat[Convergence Iterations]{%
    \includegraphics[width=0.5\linewidth]{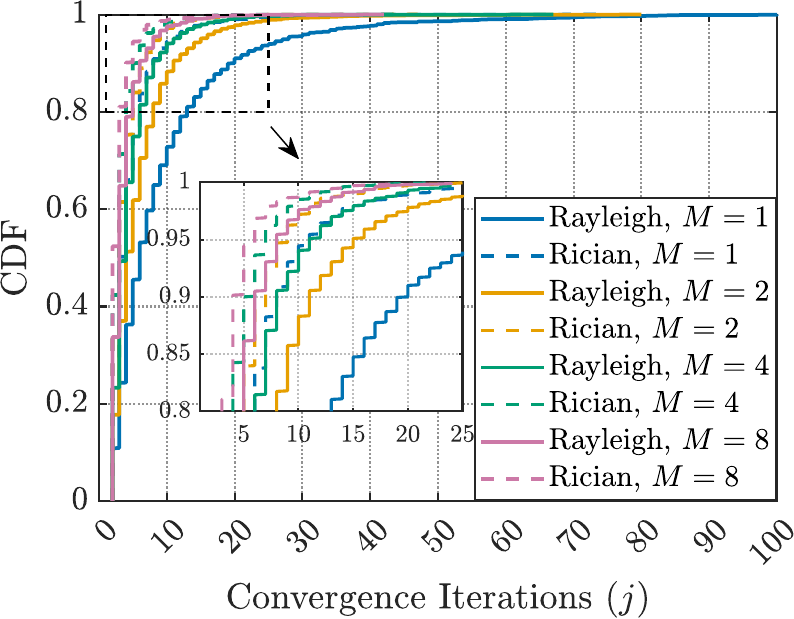}%
    \label{fig3:cdf_errors}}
  \caption{%
   Localization performance under Rayleigh and Rician fadings for different numbers of antennas $M$: 
(a) Average absolute localization error and $90$\% CI at convergence
vs. $M$ and 
(b) CDF of the number of iterations until convergence for \( M \in \{1, 2, 4, 8\} \), 
}
  \label{fig:cdf_subfigs}
\end{figure}

Fig.~\ref{fig:cdf_subfigs} 
shows how the number of antennas $M$ per subnetwork affects localization accuracy and convergence under Rayleigh and Rician fading
using the average absolute localization error at algorithm convergence as metric. As expected, the Rayleigh fading leads to higher errors due to the absence of a strong LoS path--e.g., at $M=1$, the average error is $15$\,m (Rayleigh) vs. $5$\,m (Rician). Increasing $M$ mitigates small-scale fading via spatial diversity: at $M=8$, errors drop to $3.5$\,m (Rayleigh) and $1.5$\,m (Rician), reaching $1.2$\,m and $0.6$\,m at $M=64$.


Fig.~\ref{fig3:cdf_errors} shows the cumulative distribution function (CDF) of the 
iterations \( j \) required by the algorithm to converge 
for $M \in \{1,2,4,8\}$. 
Increasing \(M\) accelerates convergence, with Rician fading converging faster due to greater channel stability. 
These results underscore the role of spatial diversity in 
accelerating convergence, and reducing localization error.

\begin{figure}[t]
  \centering
  \subfloat[Average  throughput loss]{%
    \includegraphics[width=0.475\linewidth]{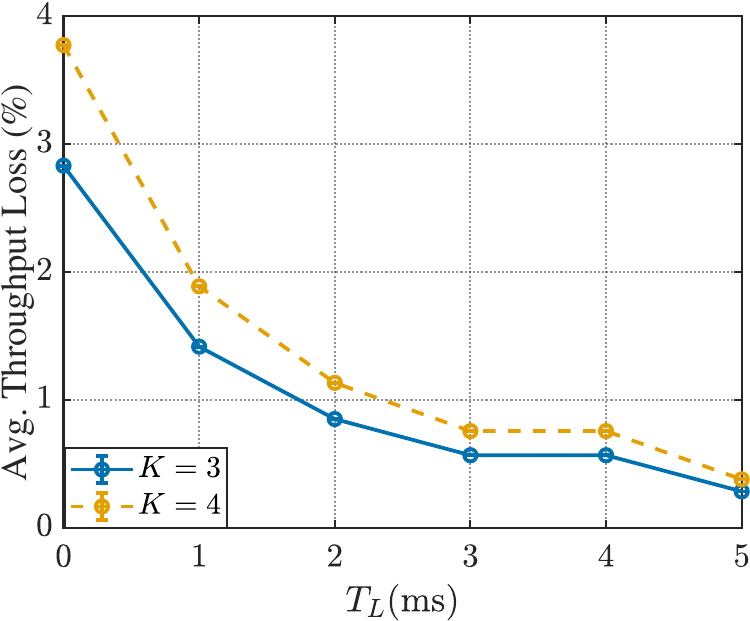}
    \label{fig4:th_loss}}
  \hfill
  \subfloat[Average localization error ]{%
    \includegraphics[width=0.495\linewidth]{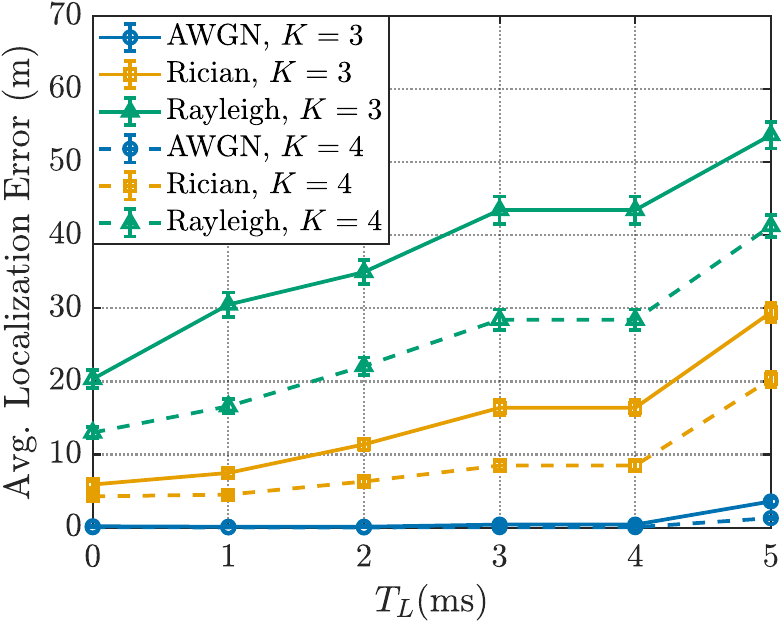}%
    \label{fig4:ac_errors}}
  \caption{ISAC trade-off and 90\% CI for 
  $T_L \in [0,5]$ ms:
    (a) Average throughput loss and 
    (b) Average localization error under AWGN, Rician 
    and Rayleigh.
    }
  \label{fig4:isac}
\end{figure}

Finally, Fig.~\ref{fig4:isac} illustrates the ISAC trade-off comparing communication throughput loss of \( 
K/(F_S\,N_{\mathrm{RB}}) \) 
and sensing accuracy as a function of the localization-feedback period $T_L$, with $K\in\{3,4\}$.
Here, the time budget is fixed as \( T_c = T\). 
As shown in Fig.~\ref{fig4:th_loss}, the average throughput loss decreases monotonically with 
$T_L$ for both values of $K$, due to the reduced 
number of sensing iterations that can be accommodated within $T_c$. 
Following~\eqref{eq:frame_constraint}, with $T_s = 1$ ms, up to 10 sensing iterations can be executed when $T_L = 0$ ms, whereas only a single iteration is feasible when $T_L = 5$ ms, leading to 
increased communication resources and 
lower throughput degradation. 
Note that for $T_L \in [3,4]$, $J_{\max}$ remains the same and so the communication and sensing performance. 
Reducing $K$ from four to three decreases throughput loss more noticeably at low $T_L$, where more sensing iterations occur (e.g., from 3.7\% to 2.8\% at $T_L=0$\,ms). At higher $T_L$, the difference diminishes (e.g., from 0.37\% to 0.28\% at $T_L=5$\,ms).


 Conversely, Fig.~\ref{fig4:ac_errors} shows the localization error grows with $T_L$ in all cases due to the fewer sensing occurrences. AWGN yields the highest accuracy, followed by Rician and Rayleigh, due to the channel randomness. Adding a fourth subnetwork improves accuracy by increasing spatial diversity, especially in fading environments. At $T_L=0$\,ms, error drops from 5.9\,m to 4.2\,m (29\%) and from 20.3\,m to 13\,m (36\%), for Rician and Rayleigh, respectively. At $T_L=5$\,ms, gains remain notable  as Rician error falls from 29\,m to 20\,m (31\%) and Rayleigh from 54\,m to 41\,m (24\%). In contrast, under AWGN channels with $T_L\le2$\,ms, adding subnetworks offers minimal sensing gains but further degrades communication. Some improvements with $K=4$ are instead visible for $T_L \geq 3$\,ms, e.g., at $T_L = 5$\,ms, the error drops from approx. 3.6~m to 1.2~m.

\section{Conclusions and Future Work}  \label{Conclusion}
We proposed a cooperative localization and iterative node selection algorithm for ISAC-enabled subnetworks. It significantly outperformed existing benchmarks, achieving sub-7\,cm localization errors in AWGN channels within only three iterations—reducing error by over 97\% compared to the best-performing benchmark under equal resource constraints.
Increasing the antenna count from 1 to 64 improved convergence 
and robustness in fading environments, underscoring the benefit of spatial diversity. 
While increasing the number of sensing subnetworks from 3 to 4 and increasing sensing iterations enhanced accuracy and mitigated fading, it also 
reduced throughput, revealing the sensing–communication trade-off.
Future work will focus on enhancements and alternative ranging mechanisms to achieve the indoor localization target of 10 cm in fading channels.


\section*{Appendix: Computation of CRLB and \textnormal{\text{e}CRLB}}
\addcontentsline{toc}{section}{Appendix}  
\label{Appendix}
Due to the unknown instantaneous channel state information, the received amplitude is simplified as a single-antenna path-loss-dependent coefficient for CRLB derivation, modeled as
\( a_n = h_n^{(m)}[r]x_n[r] = \sqrt{p_n G_t G_r \sigma_{\mathrm{rcs}} \lambda^2 / (4\pi)^3 d^4_n} \). 
Given \( R \) i.i.d. observations \( y_n^{(m)}[1], \dots, y_n^{(m)}[R] \sim \mathcal{N}(a_n, \sigma_n^2) \), the 
likelihood function for the unknown range \( d_n \) is given by
\begin{align}
L(d_n) 
&= \prod_{r=1}^{R} \frac{1}{\sqrt{2\pi\sigma_n^2}} \exp\left( -\frac{(y_n^{(m)}[r] - a_n)^2}{2\sigma_n^2} \right).
\label{eq:3}
\end{align}
The Fisher information 
is \( I(d_n) = \mathbb{E}\left[\left(\frac{\partial}{\partial d_n}\ln L(d_n)\right)^2\right]\). By taking the logarithm of~\eqref{eq:3}, 
applying the chain rule, recalling that \( z_n^{(m)}[r] = y_n^{(m)}[r] - a_n \sim \mathcal{N}(0, \sigma_n^2) \), and noting that \( \frac{\partial a_n}{\partial d_n} = -\frac{2a_n}{d_n} \), we obtain
\begin{IEEEeqnarray}{C}
I(d_n) 
= \left( \frac{2a_n}{d_n \sigma_n^2} \right)^2 \mathbb{E}\! \left[ \left( \sum_{r=1}^{R} y_n^{(m)}[r] - a_n \right)^2 \right]\! = \frac{4a_n^2 R}{d_n^2 \sigma_n^2}.
\label{eq:Fisher}\IEEEeqnarraynumspace
\end{IEEEeqnarray}

The CRLB states that the variance of any unbiased estimator is lower-bounded by the inverse of the Fisher information, i.e., 
\begin{equation}
\mathrm{Var}(\hat{d}_n) \geq \frac{d^2_n \sigma^2_n}{4a^2_n R} 
= \frac{d^6_n \sigma^2_n (4\pi)^3}{4R p_n G_t G_r \sigma_{\mathrm{rcs}} \lambda^2}.
\label{eq:10}
\end{equation}
The 
bound~\eqref{eq:10} cannot be 
evaluated, as it depends on the unknown 
target range. Moreover, its estimation through direct measurements, e.g., through ~\eqref{RES},
would incur excessive delay and resource consumption. To address this, we approximate the true distance \( d_n \) by $\widetilde{d}_n = \| \hat{\mathbf{q}}^{} - \mathbf{s}_n \|_2$, 
computed using the current estimated target location. 
Substituting \( \widetilde{d}_n \) in~\eqref{eq:10} yields 
the approximate bound eCRLB, denoted by \( \widehat{\mathrm{CRLB}}_{n \leftarrow \hat{\mathbf{q}}} \).

\bibliographystyle{IEEEtran}  
\bibliography{refs}

\begin{thebibliography}{10}
\providecommand{\url}[1]{#1}
\csname url@samestyle\endcsname
\providecommand{\newblock}{\relax}
\providecommand{\bibinfo}[2]{#2}
\providecommand{\BIBentrySTDinterwordspacing}{\spaceskip=0pt\relax}
\providecommand{\BIBentryALTinterwordstretchfactor}{4}
\providecommand{\BIBentryALTinterwordspacing}{\spaceskip=\fontdimen2\font plus
\BIBentryALTinterwordstretchfactor\fontdimen3\font minus \fontdimen4\font\relax}
\providecommand{\BIBforeignlanguage}[2]{{%
\expandafter\ifx\csname l@#1\endcsname\relax
\typeout{** WARNING: IEEEtran.bst: No hyphenation pattern has been}%
\typeout{** loaded for the language `#1'. Using the pattern for}%
\typeout{** the default language instead.}%
\else
\language=\csname l@#1\endcsname
\fi
#2}}
\providecommand{\BIBdecl}{\relax}
\BIBdecl

\bibitem{berardinelli2021extreme}
G.~Berardinelli \emph{et~al.}, ``Extreme communication in {6G}: Vision and challenges for `in{-X}' subnetworks,'' \emph{IEEE Open J. Commun. Soc.}, vol.~2, pp. 2516--2535, 2021.

\bibitem{lu2024integrated}
S.~Lu \emph{et~al.}, ``Integrated sensing and communications: Recent advances and ten open challenges,'' \emph{IEEE Internet Things J.}, vol.~11, no.~11, pp. 19\,094--19\,120, 2024.

\bibitem{Meng2025cooperative}
K.~Meng \emph{et~al.}, ``Cooperative {ISAC} networks: Opportunities and challenges,'' \emph{IEEE Wireless Commun.}, vol.~32, no.~3, pp. 212--219, 2025.

\bibitem{srinivasan2024multi}
A.~Srinivasan, U.~Singh, and O.~Tirkkonen, ``Multi-agent reinforcement learning approach scheduling for in-{X} subnetworks,'' in \emph{Proc. IEEE 100th Veh. Technol. Conf. (VTC-Fall)}, 2024, pp. 1--7.

\bibitem{hakimi2025resilient}
S.~Hakimi, R.~Adeogun, and G.~Berardinelli, ``Resilient {DNN} for joint sub-band allocation and power control in mobile factory subnetworks,'' \emph{EURASIP J. Wireless Commun. Netw.}, vol. 2025, no.~1, p.~49, 2025.

\bibitem{jiang2023collaborative}
W.~Jiang \emph{et~al.}, ``Collaborative precoding design for adjacent integrated sensing and communication base stations,'' \emph{IEEE Internet Things J.}, vol.~11, no.~9, pp. 15\,059--15\,074, 2023.

\bibitem{meng2024network}
K.~Meng \emph{et~al.}, ``Network-level integrated sensing and communication: Interference management and {BS} coordination using stochastic geometry,'' \emph{IEEE Trans. Wireless Commun.}, 2024.

\bibitem{Li2025cooperative}
H.~Li \emph{et~al.}, ``Multi-node multi-band cooperative integrated sensing and communications: State-of-the-art, challenges and opportunities,'' \emph{IEEE Wireless Commun.}, vol.~32, no.~4, pp. 180--188, 2025.

\bibitem{chen2013calculation}
C.-S. Chen \emph{et~al.}, ``Calculation of weighted geometric dilution of precision,'' \emph{J. Appl. Math.}, vol. 2013, no.~1, p. 953048, 2013.

\bibitem{liu2023snr}
R.~Liu \emph{et~al.}, ``{SNR/CRB}-constrained joint beamforming and reflection designs for {RIS-ISAC} systems,'' \emph{IEEE Trans. Wireless Commun.}, vol.~23, no.~7, pp. 7456--7470, 2023.

\bibitem{bjorck2024numerical}
{\AA}.~Bj{\"o}rck, \emph{Numerical methods for least squares problems}.\hskip 1em plus 0.5em minus 0.4em\relax SIAM, 2024.

\bibitem{luo2024channel}
C.~Luo \emph{et~al.}, ``Channel modeling framework for both communications and bistatic sensing under {3GPP} standard,'' \emph{IEEE J. Sel. Areas Sensors}, 2024.

\bibitem{zafari2019survey}
F.~Zafari, A.~Gkelias, and K.~K. Leung, ``A survey of indoor localization systems and technologies,'' \emph{IEEE Commun. Surveys Tuts.}, vol.~21, no.~3, pp. 2568--2599, 2019.

\end{thebibliography}

\end{document}